\documentclass[useAMS,usenatbib,usegraphicx,onecolumn]{mn2e}
\usepackage[normalem]{ulem}

\title[Young embedded clusters at high Galactic latitude?]{On the existence of young embedded clusters at high Galactic latitude}
\author[Turner, Carraro, \& Panko]{D.~G. Turner$^1$\thanks{Email: turner@ap.smu.ca}, G. Carraro$^2$, and E.~A. Panko$^3$\\ \\
$^1$Department of Astronomy and Physics, Saint Mary's University, Halifax, NS B3H 3C3, Canada \\
$^2$Dipartimento di Fisica e Astronomia, Universit\'{a} di Padova, Vicolo Osservatorio 3, I-35122, Padova, Italy\\
$^3$Theoretical Physics and Astronomy Department, I.~I.~Mechnikov Odessa National University,
Paster str, 42, Odessa, Ukraine
}
\begin{document}
\pagerange{\pageref{1}--\pageref{7}} \pubyear{2017}

\label{firstpage}
\maketitle

\begin{abstract}
Careful analyses of photometric and star count data available for the nine putative young clusters identified by \citet{ca15,ca16} at high Galactic latitudes reveal that none of the groups contain early-type stars, and most are not significant density enhancements above field level. 2MASS colours for stars in the groups match those of unreddened late-type dwarfs and giants, as expected for contamination by (mostly) thin disk objects. A simulation of one such field using only typical high latitude foreground stars yields a colour-magnitude diagram that is very similar to those constructed by \citet{ca15,ca16} as evidence for their young groups as well as the means of deriving their reddenings and distances. Although some of the fields are coincident with clusters of galaxies, one must conclude that there is {\it no} evidence that the putative clusters are extremely young stellar groups.
\end{abstract}


\section{Introduction}
In two recently published studies, \citet{ca15,ca16} report the discovery of nine purported extremely young open clusters at high Galactic latitude in some cases associated with suspected high-velocity hydrogen clouds in the halo. In each case the star clusters were detected by their association with hot spots in the WISE survey \citep{wr10}, and subsequently delineated by star counts. Distances and extinctions for the stellar groups were obtained from analysis of {\it JHK$_s$} photometry \citep{cu03} for group stars in the 2MASS infrared survey \citep{sk06}. The extinction and age of each group were established solely from rudimentary fitting of model isochrones for young stars to cleaned {\it JHK$_s$} colour-magnitude plots, without recourse to independent estimates of reddening for group stars. All nine clusters were found to have ages of a few million years, to be associated with high-latitude H~I clouds in a few cases, and to lie at distances of 4--7 kpc with visual extinctions $A_V$ of 0.7--1.5 magnitude, despite the fact that clusters of distant galaxies are visible in sky survey images for several of the fields. Also curious is the relative absence of hydrogen-burning main-sequence stars in the nine clusters.

The method of colour-magnitude diagram (CMD) analysis used by \citet{ca15,ca16} does not follow standard precepts that date from the 1950s and 1960s, in which establishing the nature of the reddening in cluster fields is an important first step  \citep[e.g.,][]{tu96}. That is relatively straightforward for 2MASS photometry \citep{tu11}, particularly for young B-type stars of the type postulated to populate the nine groups studied by \citet{ca15,ca16}, but was neglected in their studies. Their observational CMDs also use {\it J--K}$_{\rm s}$ instead of {\it J--H} as the temperature index, even though uncertainties in 2MASS {\it K}$_{\rm s}$-band photometry typically reach $\pm0.2$ or more near the survey limits, with uncertainties in {\it J--K}$_{\rm s}$ colour reaching values ranging from about $\pm0.25$ to perhaps $\pm0.5$, consistent with the extent of the scatter in the observed CMDs. Such large photometric scatter makes model isochrone fitting from 2MASS data a rather daunting challenge for the groups identified by \citet{ca15,ca16}.

The situation for {\it J--H} colours is typically better, although uncertainties here also reach $\pm0.20$ near the survey limits. By comparison, typical uncertainties in {\it B--V} colours from photographic photometry during the photographic plate era were as much as four times smaller, on average \citep{tw89}. It was from such considerations that {\it J--H} was advocated as the colour index most suitable for studies of open clusters from 2MASS photometry \citep{tu11}. Independent estimates of reddening are also available for all fields, given their high Galactic latitudes and in some cases noted associations with high-latitude H~I clouds, since each field contains numerous faint galaxies in the NASA/IPAC Extragalactic Database (NED, https://ned.ipac.caltech.edu), for which reddenings are cited from H~I column densities via the \citet{bh82} relations.

As demonstrated here, when the groups delineated by \citet{ca15,ca16} are studied with proper attention given to all sources of uncertainty in such studies, an inescapable conclusion is that very few of the groups are obvious star clusters. Most are simple clumpings of field stars, or possibly dissolved older stellar groups. In some cases they are projected against known clusters of distant galaxies. Most can be rejected as star clusters on the basis of star counts, although the colour and brightness data also provide useful insights into the true nature of the putative groups. Simulations of what one should expect to find for typical halo fields yield CMDs that are remarkably similar to those published by \citet{ca15,ca16}.
 
\section{Observational Data Available for the Putative Clusters}
A curious aspect of the original studies by \citet{ca15,ca16} is the manner in which star counts were made for each field, then plotted logarithmically, which only emphasizes counts for the inner group regions, which are associated with the largest uncertainties. Star counts are useful for establishing the number of likely cluster members relative to field stars observed along the same line of sight, and are presumably a necessary step in the CMD ``cleaning'' process used by \citet{ca15,ca16}. The standard method of star counts is described in detail by \citet{vb60}, as used, for example, in a recent study of the Carina cluster SAI~113 \citep{ca17}. That entailed counts made from the 2MASS data sets \citep{cu03}, which appear to have a photometric limit near $J=17$ in all fields. Such counts can guide the photometric analysis, since there cannot be {\it more} cluster members than indicated by star counts. Yet that seems to be the case in most of the fields studied by \citet{ca15,ca16}.

The present analysis therefore begins with a reanalysis of star counts for each of the fields identified by \citet{ca15,ca16}, using ring counts about the adopted cluster centres to the {\it J-}photometry limits of the 2MASS database \citep{cu03}. Data for stars in the inner ($r< 5\arcmin$) region of each field, as well as in an equal-area ring ($5\arcmin \le r < 7.07\arcmin$) surrounding the inner ring, were then analyzed using the methodology of \citet{tu11}.

For the photometric analysis we also used stars in the very young cluster Stock~16 \citep{tu85} reddened by the amounts for each cluster specified by \citet{ca15,ca16} and adjusted to the distances they found. Stock~16 is ideal for such purpose, since it contains young stars approaching the age of the putative groups identified by \citet{ca15,ca16}, with only a very small amount of differential reddening and extinction affecting the observations \citep{tu85,tu96,ca17}. Any observed scatter in the 2MASS photometry for Stock~16 stars is therefore diagnostic of what one should expect when analyzing the data with respect to model isochrones. The methodology used for stars in the 2MASS survey is that of \citet{tu11}, and Table~\ref{tab1} summarizes the results of the analyses.

The analysis for each field included star counts (top of each figure) with their uncertainties \citep[see][]{vb60}, plotted relative to the field star level indicated by rings lying outside the adopted cluster core, as established from the point where the star counts reach field level. A {\it JHK}$_{\rm s}$ two-colour diagram (2CD, middle of each figure) indicates stars in the central regions of each group (filled circles) as well as in the outer ring (open circles), along with similar data for members of Stock~16 (gray points). The resulting colour-magnitude diagram for each field (bottom of each figure) uses the same identification system. Intrinsic relations are shown as black curves, and gray curves correspond to the parameters compiled by \citet{ca15,ca16}. Discussions for each field are given below.

\begin{figure}
\begin{center}
\includegraphics[width=6cm]{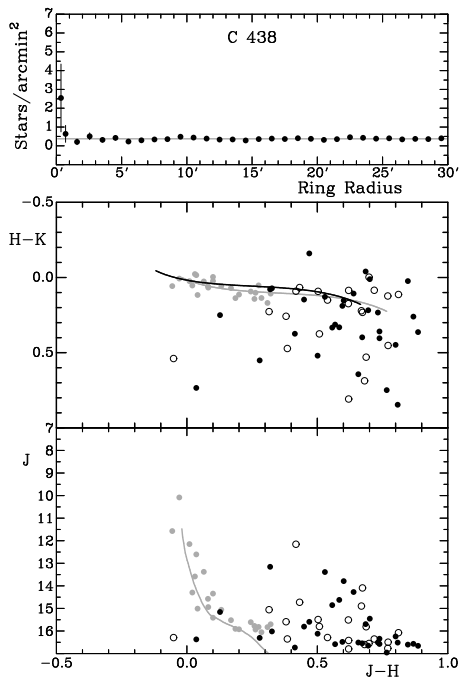}
\end{center}
\caption{Analysis of the putative group C~438 using star counts from 2MASS (upper) and {\it JHK}$_{\rm s}$ two-colour (middle, intrinsic relation plotted as black curve) and colour-magnitude (lower) diagrams. Gray lines and curves represent the adopted field star level (upper) and intrinsic relations (middle, lower) for the parameters found by \citet{ca15,ca16}. Data for Stock~16 stars (gray points) are adjusted to the same parameters.}
\label{fig1}
\end{figure}

\subsection{C~438}
Star counts in Fig.~\ref{fig1} match those of \citet{ca15}, but imply no excess of stars above field level in the putative cluster core (Table~\ref{tab1}). The implied number of cluster members, $0.8 \pm1.1$, contrasts markedly with the 23 stars in the CMD of \citet{ca15}, and the 2CD and CMD imply a lack of young stars in C~438 sharing reddenings or implied distances matching those derived by \citet{ca15}. A few potentially matching stars at the survey limits can be discounted because of their extremely uncertain colours, given the large spread in {\it J--H} values near $J\simeq 17$. C~438 lies in the line of sight to a variety of distant galaxies in NED, with GALEXASC J001914.75--184751.6 from the {\it Galaxy Evolution Explorer All-Sky Survey Source Catalog} being the closest \citep{ma90}. The implied reddening via the \citet{bh82} method is $E_{B-V}\simeq0.02$ according to NED, consistent with the very small reddenings evident in the middle section of Fig.~\ref{fig1}. The majority of objects in this field have colours and brightnesses similar to those of unreddened late-type dwarfs typical of the Galactic thin disk. C~438 is {\it not} a true cluster.

\begin{figure}
\begin{center}
\includegraphics[width=6cm]{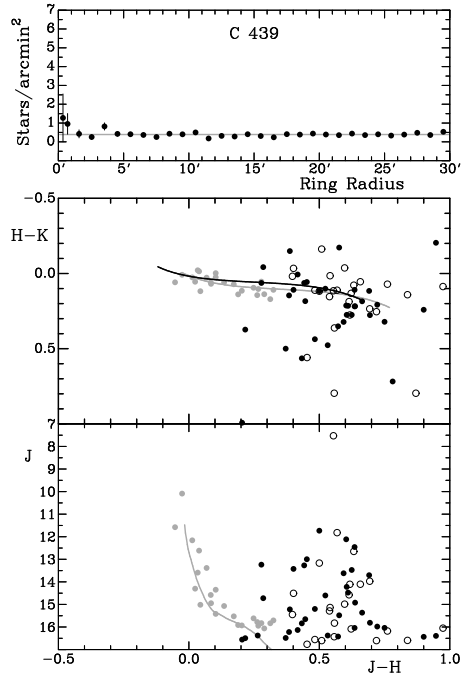}
\end{center}
\caption{Analysis of the putative group C~439 as in Fig.~\ref{fig1} using identical symbols.}
\label{fig2}
\end{figure}

\subsection{C~439}
Star counts in Fig.~\ref{fig2} and Table~\ref{tab1} also match those of \citet{ca15}, but imply essentially no excess of stars above field level in C~439, except for a few faint stars near the 2MASS survey limits that are part of the large colour spread for faint stars seen also in Fig.~\ref{fig1}. The implied number of cluster members, $1.8 \pm1.1$, again contrasts markedly with the 25 stars in the CMD of \citet{ca15}. C~439 lies in a sparse field of NED, with MRSS 539--111427 from the {\it Muenster Red Sky Survey} \citep{un03} being closest to its supposed centre. The implied reddening via the \citet{bh82} method is $E_{B-V}\simeq0.02$, as in C~438, consistent with the very small reddenings evident in Fig.~\ref{fig2}. As for C~438, the majority of objects in the field have colours and brightnesses similar to those of unreddened late-type dwarfs typical of the Galactic thin disk. C~439 appears to represent merely a statistically-insignificant blip in the general halo field.

\begin{figure}
\begin{center}
\includegraphics[width=6cm]{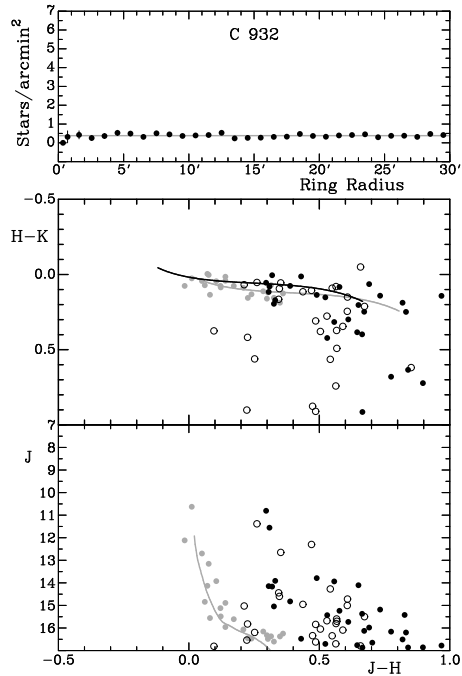}
\end{center}
\caption{Analysis of the putative group C~932 as in Fig.~\ref{fig1} using identical symbols.}
\label{fig3}
\end{figure}

\subsection{C~932}
Star counts in Fig.~\ref{fig3} for C~932 differ substantially from those given by \citet{ca16} and indicate {\it no} excess of stars above field level at the cited co-ordinates for the group \citep{ca16}: {$-0.2\pm1.1$ versus the 31 stars in their CMD. The majority of objects in the field have the colours and brightnesses expected for unreddened late-type dwarfs typical of the Galactic thin disk. The field of C~932 sits near the centre of the galaxy cluster 0211--1826 of \citet[][see Fig.~\ref{fig4}]{pf06}, with the galaxy MRSS 544--071744 lying {\bf spatially} closest to its centre \citep{un03}. NED implies a reddening of $E_{B-V}\simeq0.01$ for the field from the \citet{bh82} formula, consistent with the 2MASS colours for objects in the field, likely a mix of old stars and a few galaxies. There is no star cluster here.

\begin{figure}
\begin{center}
\includegraphics[width=6cm]{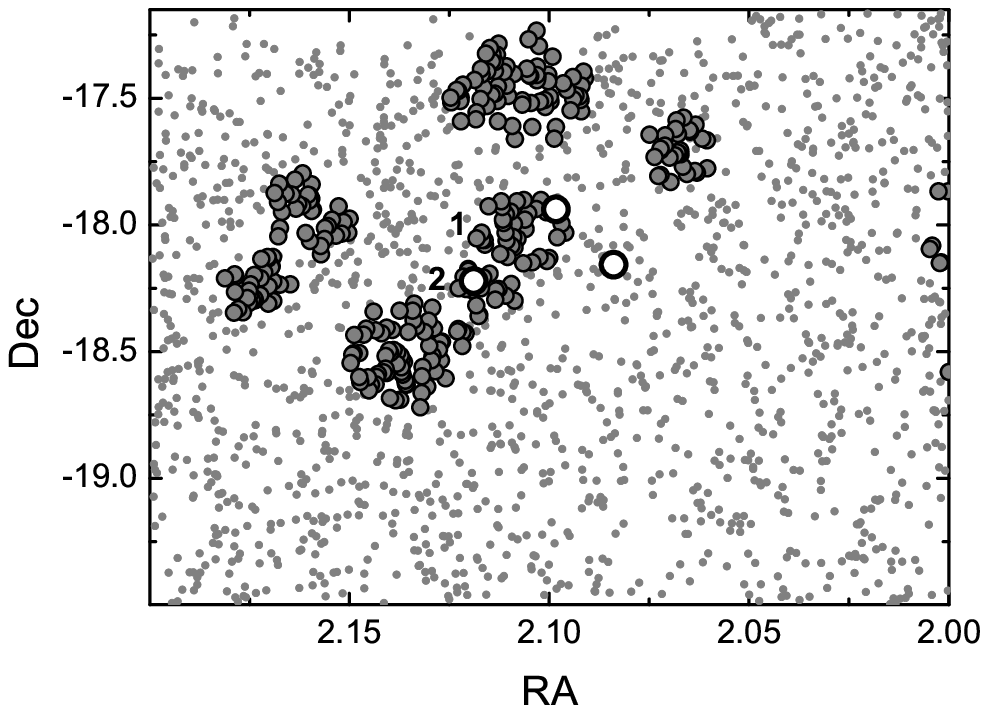}
\end{center}
\caption{Galaxies (small gray points) in the field of the putative groups C~932 (left circle), C~934 (centre circle), and C~939 (right circle). Galaxies associated with seven galaxy clusters of \citet{pf06} are highlighted as larger gray points, with clusters 0210--1802 and 0211--1826 numbered as 1 and 2, respectively.}
\label{fig4}
\end{figure}

\begin{figure}
\begin{center}
\includegraphics[width=6cm]{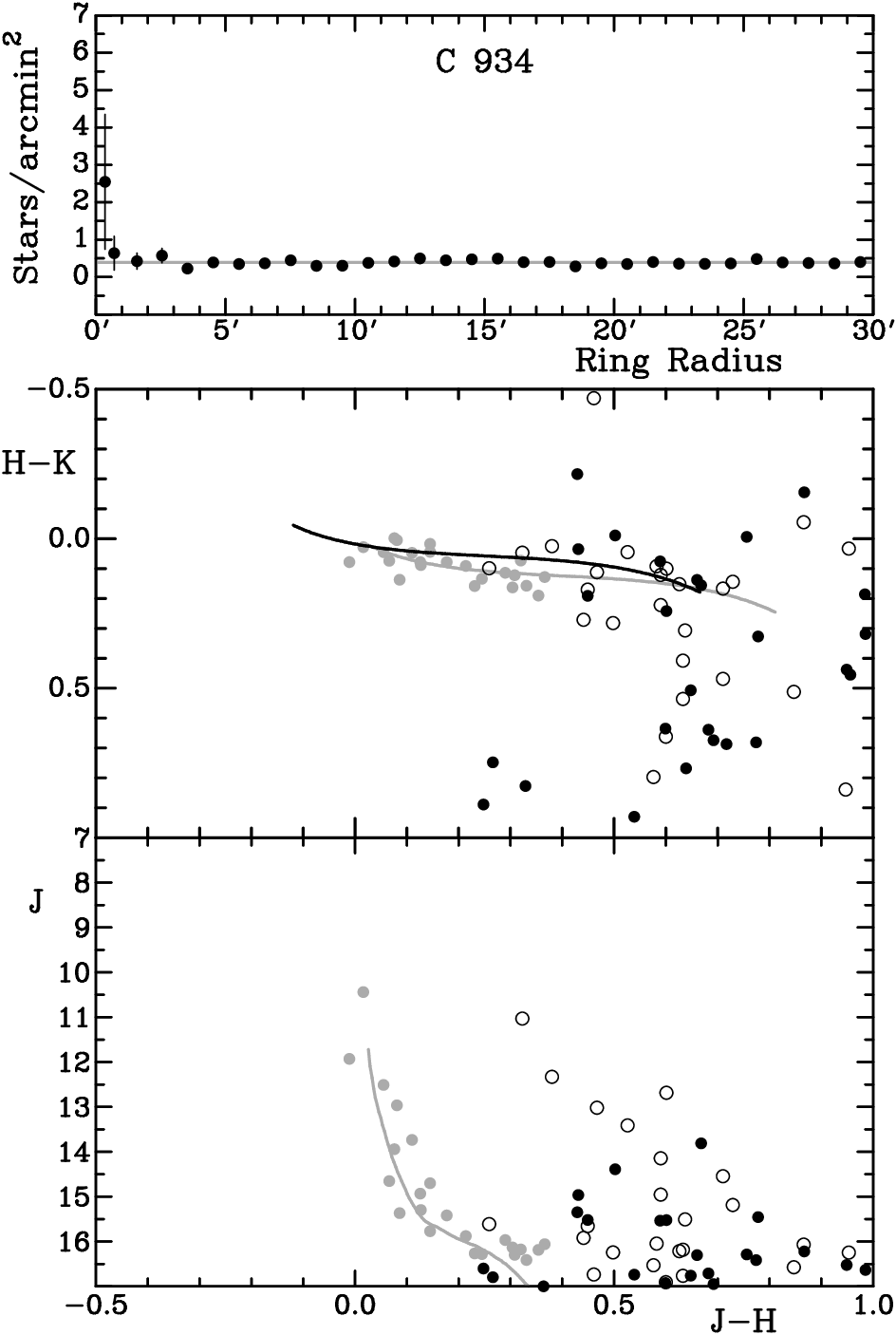}
\end{center}
\caption{Analysis of the putative group C~934 as in Fig.~\ref{fig1} using identical symbols.}
\label{fig5}
\end{figure}

\subsection{C~934}
The star counts in Fig.~\ref{fig5} for C~934 are consistent with those given by \citet{ca16}, but imply no excess of stars above field level at the cited co-ordinates: $0.8\pm1.1$ versus the 21 stars in their CMD. The majority of objects in the field have colours and brightnesses typical of unreddened late-type dwarfs of the Galactic thin disk, which constitute the dominant objects lying in directions away from the Galactic plane. C~934 lies on the northern edge of galaxy cluster 0210--1802 of \citet[][see Fig.~\ref{fig4}]{pf06}, with the galaxy closest to its cited location being APMUKS(BJ) B020331.79--181032.3 of the {\it Automated Plate Measurement United Kingdom Schmidt (B\_J\_)} survey \citep[see][]{ma90}. The reddening from NED and the \citet{bh82} formula is $E_{B-V}\simeq0.01$, as also found for C~932. The 2MASS colours suggest a similar value of $E_{B-V}\simeq0.00$. In any case, the various observations suggest that there is {\it no} cluster here.

\begin{figure}
\begin{center}
\includegraphics[width=6cm]{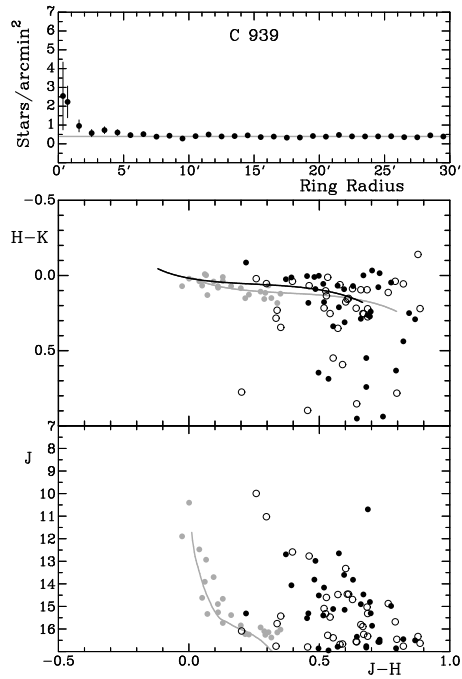}
\end{center}
\caption{Analysis of the putative group C~939 as in Fig.~\ref{fig1} using identical symbols.}
\label{fig6}
\end{figure}

\subsection{C~939}
C~939 appears to lie outside any galaxy clusters according to Fig.~\ref{fig4}, yet star counts in Fig.~\ref{fig6} do imply an excess of stars above field level within 7\arcmin\ of the cited co-ordinates for C~939 \citep{ca16}, $33.7\pm9.2$, so a poor star cluster may exist here. But the majority of objects in the field have the colours and brightnesses expected of unreddened late-type dwarfs of the Galactic thin disk, not early-type pre-main-sequence objects of the type matching the 37 stars in the CMD of \citet{ca16}. The closest galaxy to the cited co-ordinates for C~439 is GALEXASC J020707.92--181314.1, with a reddening from NED and the \citet{bh82} formula of $E_{B-V}\simeq0.01$, consistent with 2MASS colours of objects in the field (Fig.~\ref{fig6}). If C~439 is a true cluster, it must be represented by the 25 or so late-type stars in Fig.~\ref{fig6} (bottom) that appear to describe the loose main sequence of an unreddened old moving group or globular cluster, since there are clearly no early-type stars in the field. In either case the data do {\it not} match the parameters derived by \citet{ca16} for C~939. 

\begin{figure}
\begin{center}
\includegraphics[width=6cm]{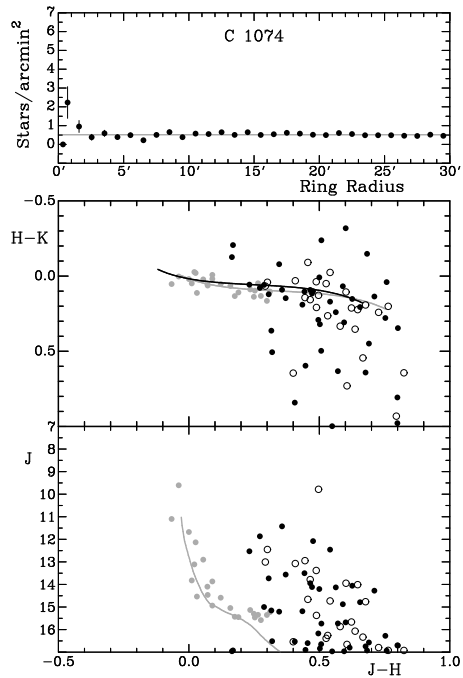}
\end{center}
\caption{Analysis of the putative group C~1074 as in Fig.~\ref{fig1} using identical symbols.}
\label{fig7}
\end{figure}

\subsection{C~1074}
There is some ambiguity in the star counts for C~1074 in Fig.~\ref{fig7} about whether or not there is an excess at the cited co-ordinates for the group. The counts are comparable to those of \citet{ca16}, but with an innermost ring omitted by them. The colours and brightnesses of group stars are those of unreddened late-type dwarfs typical of halo fields dominated by the Galactic thin disk, {\it not} young early-type stars. The implied number of cluster members, $9.5\pm2.6$, also contrasts with the 21 stars in the CMD of \citet{ca16}. The galaxy closest to the cited co-ordinates for C~1074 is SDSS J103926.90--020042.8 from the {\it Sloan Digital Sky Survey}. Its reddening from NED and the \citet{bh82} formula is $E_{B-V}\simeq0.03$, as also supported by 2MASS colours of objects in the field (Fig.~\ref{fig7}). C~1074 seems unlikely to be a real group, but its CMD does bear some similarities to that for C~939, so, like the latter, it may represent the dispersed remains of an old stellar group.

\begin{figure}
\begin{center}
\includegraphics[width=6cm]{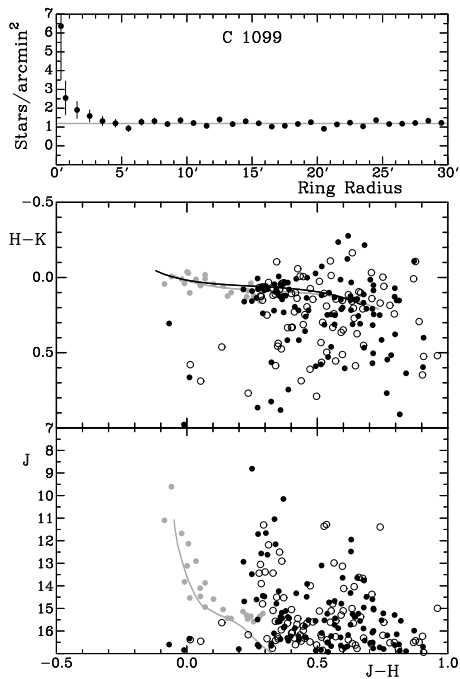}
\end{center}
\caption{Analysis of the putative group C~1099 as in Fig.~\ref{fig1} using identical symbols.}
\label{fig8}
\end{figure}

\subsection{C~1099}
Star counts for C~1099 in Fig.~\ref{fig8} are consistent with those of \citet{ca16} and indicate a slight excess of stars above field level within 4\arcmin\ of the group's cited co-ordinates:  $20.3\pm7.8$ implied members, versus 43 in their CMD. As in the similar case for C~939, the majority of objects in the field have colours and brightnesses like those of unreddened late-type dwarfs typical of halo fields dominated by thin disk stars. The object closest to the cited co-ordinates for C~1099 is 2MASX J11495819--324513 \citep{sk06}, and galaxies in this field in NED are reddened by $E_{B-V}\simeq0.08$ according to the \citet{bh82} formula. The field lies closer to the Galactic plane than others examined so far, so a larger dust extinction is expected. 2MASS colours of objects in the field (Fig.~\ref{fig8}) are consistent with a slightly larger reddening, and some of the details in the Fig.~\ref{fig8} CMD resemble those for C~939 but with a somewhat bluer main sequence. The data therefore suggest the possible existence of the remains of an old stellar group in the field, but certainly not  a group with the parameters derived by \citet{ca16}. Note, in particular, the full magnitude spread in {\it J--H} colours for stars in the field near the 2MASS limits ($J \simeq 17$).

\begin{figure}
\begin{center}
\includegraphics[width=6cm]{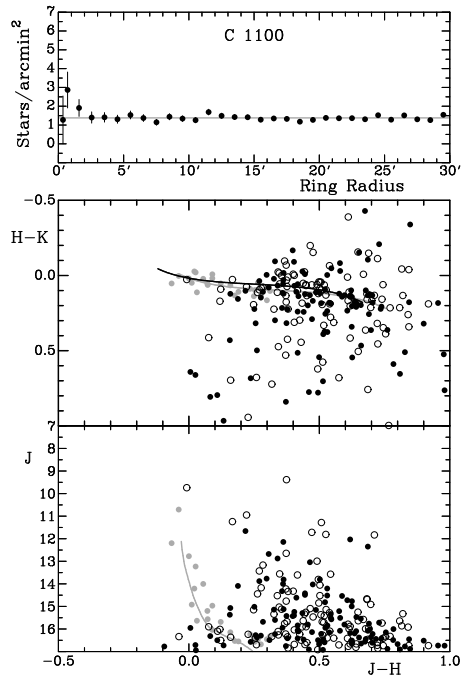}
\end{center}
\caption{Analysis of the putative group C~1100 as in Fig.~\ref{fig1} using identical symbols.}
\label{fig9}
\end{figure}

\subsection{C~1100}
There is marginal evidence for a slight excess of objects in the star counts for C~1100 in Fig.~\ref{fig9}, consistent with the counts of \citet{ca16}, but once again the stars that seem to match the parameters derived by them for the group are at the 2MASS limits and almost certainly affected by extremely large uncertainties in colour. Note, as for C~1099, the full magnitude spread in {\it J--H} colours for stars near $J \simeq 17$. The star counts also predict only $9.7\pm4.2$ members, not 51 as in the CMD of \citet{ca16}. No galaxies in NED are coincident with the core of the putative cluster, but those close to it are reddened by $E_{B-V}\simeq0.07$ according to the \citet{bh82} formula. The 2MASS colours for objects in C~1100 (Fig.~\ref{fig9}) support that, the majority having colours and brightnesses like those of unreddened late-type dwarfs typical of the Galactic thin disk. It seems unlikely that C~1100 represents a true cluster, given that the CMD of Fig.~\ref{fig9} (bottom) lacks an identifiable group of late-type main-sequence stars, as in C~939 and C~1099.

\begin{figure}
\begin{center}
\includegraphics[width=6cm]{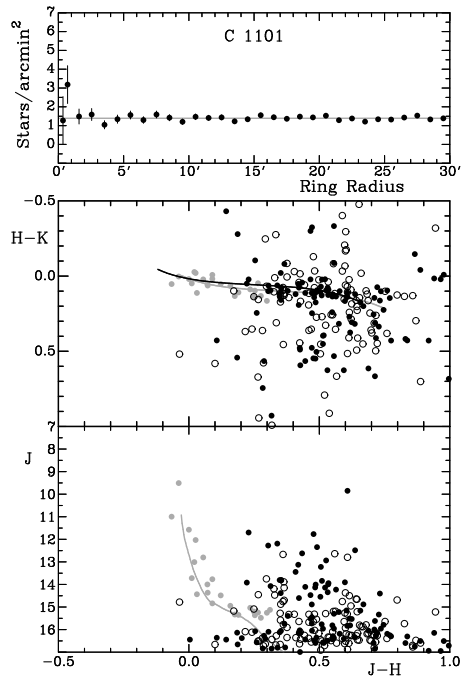}
\end{center}
\caption{Analysis of the putative group C~1101 as in Fig.~\ref{fig1} using identical symbols.}
\label{fig10}
\end{figure}

\subsection{C~1101}
Our star counts for C~1101 indicated in Fig.~\ref{fig10} are consistent with those of \citet{ca16}, but the small excess of stars within 1\arcmin\ of the cited co-ordinates,  $5.6\pm2.1$ relative to 26 in the \citet{ca16} CMD, is likely spurious. The majority of objects in the field are an excellent match to the colours and brightnesses of unreddened late-type dwarfs typical of the Galactic thin disk. Note as well the full magnitude spread in {\it J--H} colours for stars near the 2MASS limits. The object closest to the cited co-ordinates for C~1101 is 2MASX J12142207--3502149 \citep{sk06}, and NED galaxies in the field are reddened by $E_{B-V}\simeq0.07$ according to the \citet{bh82} formula. C~1101 clearly does not represent the type of young stellar group implied by \citet{ca16}.

A summary of our results for the nine putative groups of stars identified by \citet{ca15,ca16} as extremely young, embedded, open clusters is presented in Table~\ref{tab1}. In {\it no} case is there clear or unambiguous evidence for the existence of young, distant, stellar groups at the locations cited by \citet{ca15,ca16}, although in the case of C~939 and C~1099 the data do suggest the possible existence of old, unreddened groups at the designated co-ordinates. In some cases (C~932, C~934) the groups are coincident with distant clusters of galaxies, although that appears to have no effect on the star counts. Most galaxies in the {\it Muenster Red Sky Survey} are fainter than the 2MASS limits, so their influence should be minimal. The stellar population in all of the fields is otherwise consistent with that expected for regions lying away from the Galactic plane.

\setcounter{table}{0}
\begin{table}
\caption[]{Summary of analyses for Camargo et al. clusters.}
\label{tab1}
\begin{center}
\begin{tabular}{@{\extracolsep{-8pt}}ccccccc}
\hline
Group &N(inner) &N(outer) &``r$_{\rm c}$'' &N$_{\rm c}$ &Camargo &Real? \\
& & & & &CMD & \\
\hline
C~438 &30 &21 &1\arcmin &$0.8\pm1.1$ &23 &No \\
C~439 &38 &25 &1\arcmin &$1.8\pm1.1$ &25 &No \\
C~932 &32 &22 &0\arcmin &$-0.2\pm1.1$ &31 &No \\
C~934 &31 &27 &1\arcmin &$0.8\pm1.1$ &21 &No \\
C~939 &41 &36 &7\arcmin &$33.7\pm9.2$ &37 &Possibly \\
C~1074 &45 &27 &2\arcmin &$9.5\pm2.6$ &21 &Possibly \\
C~1099 &114 &86 &4\arcmin &$20.3\pm7.8$ &43 &Possibly \\
C~1100 &117 &111 &2\arcmin &$9.7\pm4.2$ &51 &No \\
C~1101 &110 &109 &1\arcmin &$5.6\pm2.1$ &26 &No \\
\hline
\end{tabular}
\end{center}
\end{table}

\section{Simulation of a Cluster Field}
In order to probe the Galactic contamination along lines of sight toward the putative clusters of \citet{ca15,ca16}, Monte Carlo simulations were performed based on the \citet{gi05} Galaxy model (TRILEGAL) in order to create synthetic CMDs of the relevant sky regions using {\it J--K}$_{\rm s}$ as the temperature index. For the sake of illustrating the results, the field of Camargo 438 (C~438) was selected as representative of a typical halo group. The Galactic model includes prescriptions at different brightness levels for the expected line-of-sight stellar contribution from the halo, bulge, and both thin and thick disks. Several trials (20) were run with TRILEGAL by varying the random seed in order to make the realizations statistically significant. The synthetic CMDs were then smeared by adding realistic photometric uncertainties typical of 2MASS photometry. Simulations were run for the line of sight coincident with the area indicated for C~438, namely for an 8\arcmin\ angular radius.

\begin{figure}
\begin{center}
\includegraphics[width=7cm]{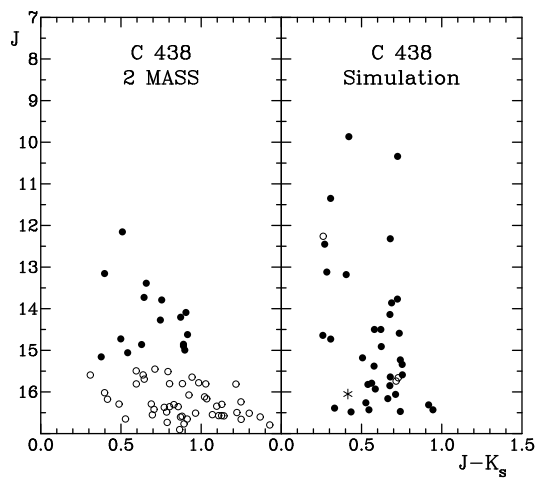}
\end{center}
\caption{Simulation of the Galactic contribution to a field of radius 8\arcmin\ at the location of C~438 (right): thin disk stars (filled circles), thick disk stars (open circles), and halo stars (star symbol). The 2MASS data for stars in the field are also shown (left): stars with {\it J-}band uncertainties no larger than $\pm0.05$ (filled circles), and stars with larger {\it J-}band uncertainties (open circles).}
\label{fig11}
\end{figure}

The results are shown in the right panel of Fig.~\ref{fig11}, where the contributions of the various Galactic components are indicated using different symbols. The greatest contribution is from thin disk stars, with only marginal contributions from the thick disk and halo. Given the halo location of C~438, no contribution is expected from bulge stars. From the inferred surface gravities of the objects in the simulations, the contribution from thin and thick disk stars is entirely from dwarfs, whereas the lone halo star is a giant. The number of stars and their distribution in the simulated CMD is entirely consistent with observations, certainly for C~438, but also for most of the other clusters, although many of the latter are located along lines of sight lying closer to the Galactic plane.

\section{Discussion}
It should be evident from the analyses of \S2 that optimum matching of observational open cluster CMDs to stellar evolutionary model predictions should initially be concerned with establishing a precise knowledge of the reddening in the field. Both foreground and differential reddening may be present, although for the halo fields examined here only foreground reddening is expected. The original \citet{ca15,ca16} studies appear to have been biased by pre-expectations based upon the fact that each field contained far-infrared emission regions in the WISE survey \citep{wr10} as well as evidence for high latitude H~I gas in several cases. But high latitude H~I gas is not uncommon. In many cases, Galactic H~I clouds are coincident with warm spots in the WMAP and PLANCK surveys \citep{vs07}, although the latter are somewhat less than 1$\degr$ in diameter, rather than the few arcminutes of the Camargo groups, and generally attributed to acoustic bubbles in the cosmic microwave background radiation \citep{hw96}. Other mechanisms may also explain their origin \citep[e.g.,][]{vs16}.

The main point, however, is that the methodology employed by \citet{ca15,ca16} leads to inconsistencies that are not supported by the star count or photometric data available. The star counts alone suggest that the method of ``cleaning'' the CMDs for contamination by field stars needs reworking. In no case were the number of potential members of the putative groups comparable to those adopted by \citet{ca15,ca16} in their group CMDs. That point is important, given that the same group has used an identical methodology to study embedded clusters in the Galactic plane.

\setcounter{table}{1}
\begin{table}
\caption[]{Star counts in the fields of putative clusters.}
\label{tab2}
\begin{center}
\begin{tabular}{@{\extracolsep{-0pt}}ccc}
\hline
Group &$|${\it b}$|$($\degr$) &Stars arcmin$^{-1}$ \\
\hline
C~438 &78.86 &0.3746 \\
C~439 &77.84 &0.3881 \\
C~444 &00.52 &5.3363 \\
C~932 &70.83 &0.3810 \\
C~934 &70.54 &0.3895 \\
C~939 &70.43 &0.3945 \\
C~1074 &46.89 &0.5165 \\
C~1099 &28.41 &1.1886 \\
C~1100 &27.41 &1.3788 \\
C~1101 &27.22 &1.3908 \\
\hline
\end{tabular}
\end{center}
\end{table}

\begin{figure}
\begin{center}
\includegraphics[width=8cm]{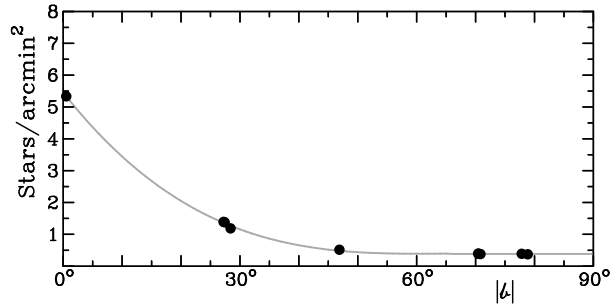}
\end{center}
\caption{Derived field star densities, to $J\simeq17$, in the fields of 10 putative Camargo clusters, including the 9 studied here, plotted as a function of absolute Galactic latitude. The gray curve represents a simple polynomial fit to the data.}
\label{fig12}
\end{figure}

\citet{ca15,ca16} might have noticed the problem of their interpretation of the high latitude groups if they had examined the field star densities in their fields. The data for the nine fields studied here, as well as for an additional Camargo et al. group lying in the Galactic plane, are summarized in Table~\ref{tab2}, with field star densities plotted in Fig.~\ref{fig12}. It should be clear that the star densities in each field are consistent with what is expected from contamination by thin disk stars, namely an exponential dropoff with increasing angular distance from the Galactic plane. There is no room for heavily-reddened stellar groups in these regions of the Galactic halo, either in the 2CDs for each field or in the star counts.

\subsection*{ACKNOWLEDGEMENTS}
This publication makes use of data products from the Two Micron All Sky Survey, which is a joint project of the University of Massachusetts and the Infrared Processing and Analysis Center/California Institute of Technology, funded by the National Aeronautics and Space Administration and the National Science Foundation.

\end{document}